\documentclass[numberedappendix,12pt,preprint]{emulateapj}

\shorttitle{utility of $H(z)$ on dark energy evolution}
\shortauthors{Meng et al. (2015)}

\usepackage{color}			      
\definecolor{midgray}{gray}{0.4}		
\definecolor{orange}{rgb}{1,0.5,0}    
\usepackage{natbib}
\usepackage[colorlinks=true,citecolor=midgray,linkcolor=midgray]{hyperref}        
\usepackage{amsmath,amssymb,amsxtra,amsfonts}
\usepackage{aas_macros}
\usepackage{graphicx}
\usepackage{epsfig}
\usepackage{txfonts}

\newcommand{\densm}{\Omega_{\mathrm{m}}}

\newcommand{\hunit}{\mathrm{km \ s^{-1} \ Mpc^{-1}}}
\newcommand{\Om} {\ensuremath{\Omega_{\rm{m}}}}
\newcommand{\lya}{Lyman-$\alpha$}

\begin{document}


\title{Utility of observational Hubble parameter data on dark energy evolution}

\author{Xiao-Lei~Meng$^{1,2}$, Xin~Wang$^2$, Shi-Yu~Li$^{1}$, Tong-Jie~Zhang$^{1}$}

\altaffiltext{1}{Department of Astronomy, Beijing Normal University, Beijing, 100875, China}
\altaffiltext{1}{Department of Physics, University of California, Santa Barbara, CA, 93106-9530, USA}
\email{tjzhang@bnu.edu.cn}

\begin{abstract}
Aiming at exploring the nature of dark energy, we use thirty-six observational Hubble parameter data (OHD) in the redshift range $0 \leqslant z \leqslant 2.36$ to make a cosmological model-independent test of the two-point $Omh^2(z_{2};z_{1})$ diagnostic. In $\Lambda$CDM, we have $Omh^2 \equiv \Om h^2$, where $\Om$ is the matter density parameter at present. We bin all the OHD into four data points to mitigate the observational contaminations. By comparing with the value of  $\Om h^2$ which is constrained tightly by the Planck observations, our results show that in all six testing pairs of $Omh^2$ there are two testing pairs are consistent with $\Lambda$CDM at $1\sigma$ confidence level (CL), whereas for another two of them $\Lambda$CDM can only be accommodated at $2\sigma$ CL. Particularly, for remaining two pairs, $\Lambda$CDM is not compatible even at $2\sigma$ CL. Therefore it is reasonable that although deviations from $\Lambda$CDM exist for some pairs, cautiously, we cannot rule out the validity of $\Lambda$CDM. We further apply two methods to derive the value of Hubble constant $H_0$ utilizing the two-point $Omh^2(z_{2};z_{1})$ diagnostic. We obtain $H_0 = 71.23\pm1.54$ $\hunit$ from inverse variance weighted $Omh^2$ value (method (I)) and $H_0 = 69.37\pm1.59$ $\hunit$ that the $Omh^2$ value originates from Planck measurement (method (II)), both at $1\sigma$ CL. Finally, we explore how the error in OHD propagate into $w(z)$ at certain redshift during the reconstruction of $w(z)$. We argue that the current precision on OHD is not sufficient small to ensure the reconstruction of $w(z)$ in an acceptable error range, especially at the low redshift.

\end{abstract}


\section{Introduction}\label{sect:intro}

In the past few decades, a number of approaches have been well established to quantitatively study the expansion history and structure 
growth of the universe \citep[see][for recent reviews]{2008ARA&A..46..385F,2013PhR...530...87W}. Originally aimed at measuring the 
cosmic deceleration rate \citep{1998AJ....116.1009R,1999ApJ...517..565P}, the observations of Type Ia supernovae (SNIa) are 
instead providing ample evidence for an accelerating expansion to an increasing precision \citep{2012ApJ...746...85S}. This 
acceleration is also strongly supported by other complementary probes, including the measurements of the baryon acoustic 
oscillation (BAO) features \citep{2005ApJ...633..560E}, the weak gravitational lensing \citep{2013MNRAS.430.2200K}, the abundance 
of galaxy clusters \citep{2013ApJ...763..147B}, the cosmic microwave background (CMB) anisotropies 
\citep{2015arXiv150201589P}, the linear growth of large-scale structure \citep{2002ApJ...572..140D}, and the 
Hubble constant $H_0$ \citep{2012ApJ...758...24F}. As the primary motive of modern cosmology shifts from whether the universe is 
accelerating to why, a robust physical model that explains the cosmic acceleration is still under debate.

A prevailing interpretation is the existence of an exotic energy constituent -- often coined dark energy (DE) -- with negative equation of state 
(EOS) parameter $w\equiv p_{\rm DE}/\rho_{\rm DE}$. By far, the most popular model for DE remains to be the simple cosmological 
constant cold dark matter model ($\Lambda \rm CDM$), with $w=-1$ at all cosmic periods \citep{2011CoTPh..56..525L}. However as popular as the $\Lambda \rm 
CDM$ model is, it still suffers from the fine tuning and coincidence problems \citep{1989RvMP...61....1W, 1999PhRvL..82..896Z}.  
In addition, it has been noticeably argued about the possibility of DE evolving its EOS, i.e. the dynamical DE models 
($w=w(z)$), amongst which there exit Quintessence \citep[$w>-1$,][]{1988PhRvD..37.3406R, 1988NuPhB.302..668W}, Phantom 
\citep[$w<-1$,][]{2002PhLB..545...23C}, K-essence \citep[$w>-1~\textrm{or}~w<-1$,][]{2000PhRvL..85.4438A, 2001PhRvD..63j3510A}, 
and especially Quintom \citep[$w$ crossing -1,][]{2005PhLB..607...35F, 2006PhLB..634..101F} models. Nonetheless, all these models 
still await more physically motivated, profound understanding. In the meantime, it is crucial to establish tests which are based 
upon direct observations and capable of unveiling, if any, dynamical features of DE. One of these diagnostics is 
$Om(z)$, which is defined as a function of redshift $z$ \citep{2008PhRvD..78j3502S,2008PhRvL.101r1301Z}, i.e.,
\begin{equation}
   \label{eq:Om}
   Om(z) = \frac{\tilde{h}^2(z)-1}{(1+z)^3-1},
\end{equation}
with $\tilde{h}=\frac{H(z)}{H_{0}}$ and $H(z)$ denoting the Hubble expansion rate. $Om(z)$ has a property of being \Om{} 
-- the matter density parameter at present -- for $\omega=-1$. Moreover, 
\citet{2012PhRvD..86j3527S} revised this diagnostic to accommodate two-point situations, i.e.,
\begin{equation}
   \label{eq:Om2points}
   Om(z_{2};z_{1}) = \frac{\tilde{h}^2(z_{2})-\tilde{h}^2(z_{1})}{(1+z_{2})^3-(1+z_{1})^3}.
\end{equation}
In this case, the measurement of $Om(z_{2};z_{1}) \equiv \Om$ is a remarkable piece of evidence for an underlying $\Lambda$CDM 
model. In other words, the measurement of $Om(z_{2};z_{1}) \ne \Om$ implies a deviation from $\Lambda$CDM and the fact 
that an evolving EOS with redshift should be considered. Consequently, according to the observations of the 
cosmic expansion history, we can successfully distinguish between DE models and perform correlative studies in 
cosmology using this diagnostic.

In this paper, we first summarize the observational methods for Hubble parameter measurements and introduce the currently available data 
sets in Section~\ref{sect:data}. In Section~\ref{sect:diaghubcons}, using the observational $H(z)$ data (OHD) hereinafter 
mentioned, we test the $\Lambda$CDM model, according to the diagnostic defined above. We also use this diagnostic to derive $H_0$ 
values in this section. In Section~\ref{sect:errorseffect}, we turn to trace out the effect of OHD errors on the reconstruction of 
DE EOS. Finally, we summarize our results and discuss limitations as well as prospects in Section~\ref{sect:conclu}.

\section{The observational $H(z)$ data sets}\label{sect:data}

OHD can be used to constrain cosmological parameters because they are obtained from model-independent direct observations.
Until now, two methods have been developed to measure OHD: galaxy differential age and radial BAO size methods 
\citep{2010AdAst2010E..81Z}.  \citet{2002ApJ...573...37J} first proposed that relative galaxy ages can be used to obtain $H(z)$ 
values and they reported one $H(z)$ measurement at $z \sim 0.1$ in their later work \citep{2003ApJ...593..622J}.  
\citet{2005PhRvD..71l3001S} added additional eight $H(z)$ points in the redshift range between 0.17 and 1.75 from differential 
ages of passively evolving galaxies, and further constrained the redshift dependence of the DE potential by reconstructing it as a 
function of redshift. \citet{2010JCAP...02..008S} provided two new determinations from red-envelope galaxies and then constrained 
cosmological parameters including curvature through the joint analysis of CMB data. Furthermore, 
\citet{2012JCAP...08..006M} obtained eight new measurements of $H(z)$ from the differential spectroscopic evolution of early-type, 
massive, red elliptical galaxies which can be used as standard cosmic chronometers. Using luminous red galaxies from Sloan 
Digital Sky Survey (SDSS) Data Release 7 (DR7), \citet{2012MNRAS.426..226C} measured a new $H(z)$ point at $z = 0.35$. Later, by 
applying the galaxy differential age method to SDSS DR7, \citet{2014RAA....14.1221Z} expanded the $H(z)$ data sample by four new points. 
Taking advantage of near-infrared spectroscopy of high redshift galaxies, \citet{2015MNRAS.450L..16M} obtained two latest 
measurements of $H(z)$. 

\begin{figure}
    \includegraphics[width=0.5\textwidth]{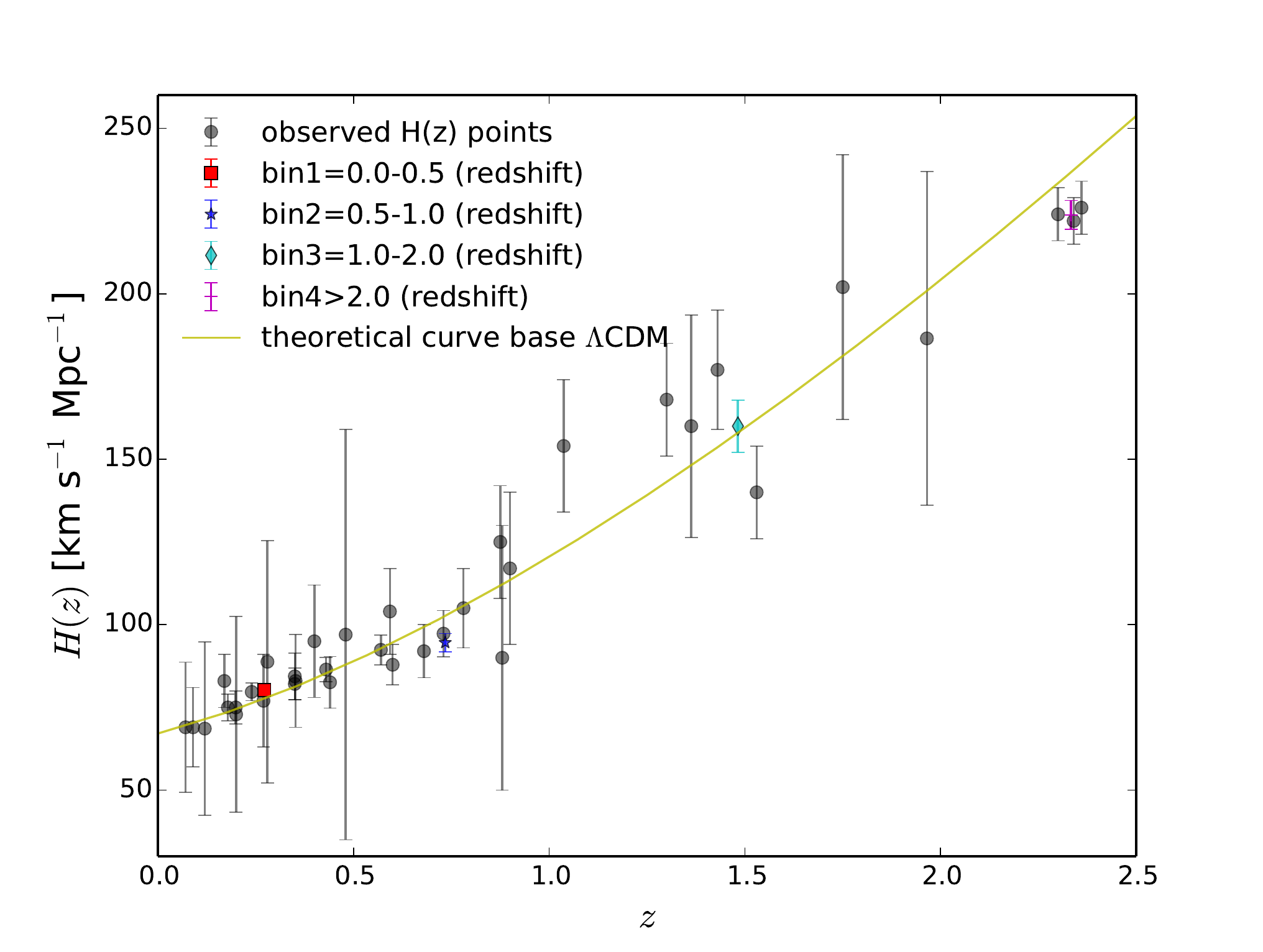}
    \caption{The full OHD set, binned $H(z)$ data and standard flat $\Lambda$CDM model. The full and binned OHD are represented by 
    gray and color-coded points, respectively. The yellow curve shows theoretical $H(z)$ evolution based upon the adopted 
    $\Lambda$CDM model which is described in Section~\ref{sect:data}.}
    \label{fig:hzobs}
\end{figure}

In addition, $H(z)$ can also be extracted from the detection of radial BAO features. \citet{2009MNRAS.399.1663G} first obtained two $H(z)$ 
data points using the BAO peak position as a standard ruler in the radial direction. \citet{2012MNRAS.425..405B} further combined the 
measurements of BAO peaks and the Alcock-Paczynski distortion to find three other $H(z)$ results. \citet{2013MNRAS.429.1514S} provided a 
$H(z)$ point at $z = 0.57$ from the BOSS DR9 CMASS sample. \citet{2013MNRAS.431.2834X} used the BAO signals from the SDSS DR7 luminous red 
galaxy sample to derive another observational $H(z)$ measurement. The $H(z)$ values determined based upon BAO features in 
the \lya forest of SDSS-III quasars were presented by \citet{2013A&A...552A..96B}, \citet{2014JCAP...05..027F}, and 
\citet{2015A&A...574A..59D}, which are the farthest observed $H(z)$ results so far. The above data points are all presented in 
Table~\ref{tab:Hzd} and marked in Figure~\ref{fig:hzobs}.

We use a $\Lambda$CDM model with no curvature term to compare theoretical values of Hubble parameter with the OHD results, with 
the Hubble parameter given by
\begin{equation}
  H(z) = H_0 \sqrt{\Om (1+z)^3+1-\Om},
\end{equation}
where cosmological parameters take values from the Planck temperature power spectrum measurements \citet{2014A&A...571A..16P}. The 
best fit value of $H_0$ is 67.11 $\hunit$, and $\densm$ is 0.3175. The theoretical computation of $H(z)$ based upon this 
$\Lambda$CDM is also shown in Figure~\ref{fig:hzobs}.

Being independent observational data, $H(z)$ determinations have been frequently used in cosmological research. One of the leading 
purposes is using them to constrain DE. \citet{2002ApJ...573...37J} first proposed that $H(z)$ measurements can be used to 
constrain DE EOS at high redshifts. \citet{2005PhRvD..71l3001S} derived constraints on DE potential using $H(z)$ results and 
supernova data. \citet{2006ApJ...650L...5S} began applying these measurements to constraining cosmological parameters in various DE 
models. In the meanwhile, DE evolution came into its own as an active research field in the last twenty years 
\citep{2000IJMPD...9..373S,2001LRR.....4....1C,2003RvMP...75..559P,2006IJMPD..15.1753C,2012PhR...513....1C}. To sum up, the OHD 
are proved to be very promising towards understanding the nature of DE.

\begin{deluxetable}{lccc}
    \tablecaption{The currently available OHD measurements\label{tab:Hzd}}
    \tablehead{
	\colhead{$z$} &
	\colhead{$H(z)$} &
	\colhead{Method} &
	\colhead{Ref.}
    }
    \startdata
    $0.0708$   &  $69.0\pm19.68$      &  I    &  Zhang et al. (2014)   \\
    $0.09$       &  $69.0\pm12.0$        &  I    &  Jimenez et al. (2003)   \\
    $0.12$       &  $68.6\pm26.2$        &  I    &  Zhang et al. (2014)   \\
    $0.17$       &  $83.0\pm8.0$          &  I    &  Simon et al. (2005)     \\
    $0.179$     &  $75.0\pm4.0$          &  I    &  Moresco et al. (2012)     \\
    $0.199$     &  $75.0\pm5.0$          &  I    &  Moresco et al. (2012)     \\
    $0.2$         &  $72.9\pm29.6$        &  I    &  Zhang et al. (2014)   \\
    $0.240$     &  $79.69\pm2.65$      &  II   &  Gaztanaga et al. (2009)   \\
    $0.27$       &  $77.0\pm14.0$        &  I    &    Simon et al. (2005)   \\
    $0.28$       &  $88.8\pm36.6$        &  I    &  Zhang et al. (2014)   \\
    $0.35$       &  $82.1^{+4.8}_{-4.9}$ & I   &  Chuang et al. (2012) \\
    $0.35$       &  $84.4\pm7.0$          &  II   &   Xu et al. (2013)  \\
    $0.352$     &  $83.0\pm14.0$        &  I    &  Moresco et al. (2012)   \\
    $0.4$         &  $95\pm17.0$           &  I    &  Simon et al. (2005)     \\
    $0.43$     &  $86.45\pm3.68$        &  II   &  Gaztanaga et al. (2009)   \\
    $0.44$       & $82.6\pm7.8$           &  II   &  Blake et al. (2012)  \\
    $0.48$       &  $97.0\pm62.0$        &  I    &  Stern et al. (2010)     \\
    $0.57$       &  $92.4\pm4.5$          &  II   &  Samushia et al. (2013)   \\
    $0.593$     &  $104.0\pm13.0$      &  I    &  Moresco et al. (2012)   \\
    $0.6$         &  $87.9\pm6.1$          &  II   &  Blake et al. (2012)   \\
    $0.68$       &  $92.0\pm8.0$          &  I    &  Moresco et al. (2012)   \\
    $0.73$       &  $97.3\pm7.0$          &  II   &  Blake et al. (2012)  \\
    $0.781$     &  $105.0\pm12.0$      &  I    &  Moresco et al. (2012)   \\
    $0.875$     &  $125.0\pm17.0$      &  I    &  Moresco et al. (2012)   \\
    $0.88$       &  $90.0\pm40.0$        &  I    &  Stern et al. (2010)     \\
    $0.9$         &  $117.0\pm23.0$      &  I    &  Simon et al. (2005)  \\
    $1.037$     &  $154.0\pm20.0$      &  I    &  Moresco et al. (2012)   \\
    $1.3$         &  $168.0\pm17.0$      &  I    &  Simon et al. (2005)     \\
    $1.363$     &  $160.0\pm33.6$      &  I    &  Moresco (2015)  \\
    $1.43$       &  $177.0\pm18.0$      &  I    &  Simon et al. (2005)     \\
    $1.53$       &  $140.0\pm14.0$      &  I    &  Simon et al. (2005)     \\
    $1.75$       &  $202.0\pm40.0$      &  I    &  Simon et al. (2005)     \\
    $1.965$     &  $186.5\pm50.4$      &  I    &   Moresco (2015)  \\
    $2.3$         &  $224.0\pm8.0$        &  II   &   Busca et al. (2013)  \\
    $2.34$       &  $222.0\pm7.0$        &  II   &  Delubac et al. (2015)   \\
    $2.36$       &  $226.0\pm8.0$        &  II   &  Font-Ribera et al. (2014) \\
    \enddata
    \tablecomments{Here the unit of $H(z)$ is $\hunit$. ``I'' quoted in this table means that the $H(z)$ value is deduced from 
    the differential age method, whereas ``II'' corresponds to that obtained from the radial BAO method.}
\end{deluxetable}

In the next section, the OHD in Table~\ref{tab:Hzd} are used to test $\Lambda$CDM models under two-point $Omh^2(z_{2};z_{1})$ diagnostic. 
Then, according to this diagnostic, we derive model-dependent values of $H_0$ assuming that $\Lambda$CDM is valid. 

\section{The applications of $Omh^2(z_{2};z_{1})$ diagnostic to OHD}\label{sect:diaghubcons}

\subsection{The test of $\Lambda$CDM}\label{ssect:testom}

In our analysis, the validity of $Om(z)$ diagnostic can be tested using $H(z)$ results from cosmological independent measurements.  
Unlike the cases where only some special $H(z)$ data from the radial BAO method are employed to test DE models 
\citep{2014ApJ...793L..40S,2014arXiv1406.7695H}, thirty-six $H(z)$ data points (as shown in Table~\ref{tab:Hzd}) are considered in 
this work. For the sake of mitigating the contaminations from systematics of different $H(z)$ observational methods performed by 
several research groups, we bin all the OHD into four data points in four redshift ranges: $0.0-0.5$, $0.5-1.0$, $1.0-2.0$, and 
$>2.0$. We take an inverse variance weighted average of all the selected data in each redshift range in the following manner. 
Assuming that $H_i(z)$ represents the $i$th observational Hubble parameter data point with $\sigma_{H_i(z)}$ denoting its reported 
observational uncertainty, in light of conventional data reduction techniques by \citet[Chap.~4]{2003drea.book.....B}, it is 
straightforward to obtain
\begin{align}\label{eq:Hzbar}
   \bar{H}(z) = & \frac{\sum_i \left(H_i(z)/\sigma^2_{H_i(z)}\right)}{\sum_i 1/\sigma^2_{H_i (z)}}, \\
   \sigma^2_{\bar{H}(z)} = & \frac{1}{\sum_i 1/\sigma^2_{H_i(z)}},
\end{align}
where $\bar{H}(z)$ stands for the weighted mean Hubble parameter in the corresponding redshift range, and $\sigma_{\bar{H}(z)}$ is 
its uncertainty. As a consequence, four weighted means of $H(z)$ and corresponding uncertainties are computed to be
$80.28\pm1.51$ $\hunit$, $94.59\pm2.76$ $\hunit$, $159.99\pm7.89$ $\hunit$, $223.81\pm4.4$ $\hunit$, at redshift $z_1=0.27$, 
$z_2=0.73$, $z_3=1.48$, $z_4=2.33$ respectively. These results are also shown in Figure \ref{fig:hzobs}.

If $Om(z_{2};z_{1})$ is always a constant at any redshifts, then it demonstrates that the DE is of the cosmological constant 
nature. 
In order to compare directly with the results from CMB, \citet{2014ApJ...793L..40S} introduced a more convenient expression of the two-point 
diagnostic, i.e.,
\begin{equation}
   \label{eq:Omh22points}
   Omh^2(z_{2};z_{1})=\frac{h^2(z_{2})-h^2(z_{1})}{(1+z_{2})^3-(1+z_{1})^3},
\end{equation}
where $h(z) = H(z)/100$ $\hunit$. The four $H(z)$ points calculated based upon the aforementioned binning method therefore 
yields six model-independent measurements of the $Omh^2(z_{2};z_{1})$ diagnostic, which are shown in Figure 2.

\begin{figure}
    \includegraphics[width=0.5\textwidth]{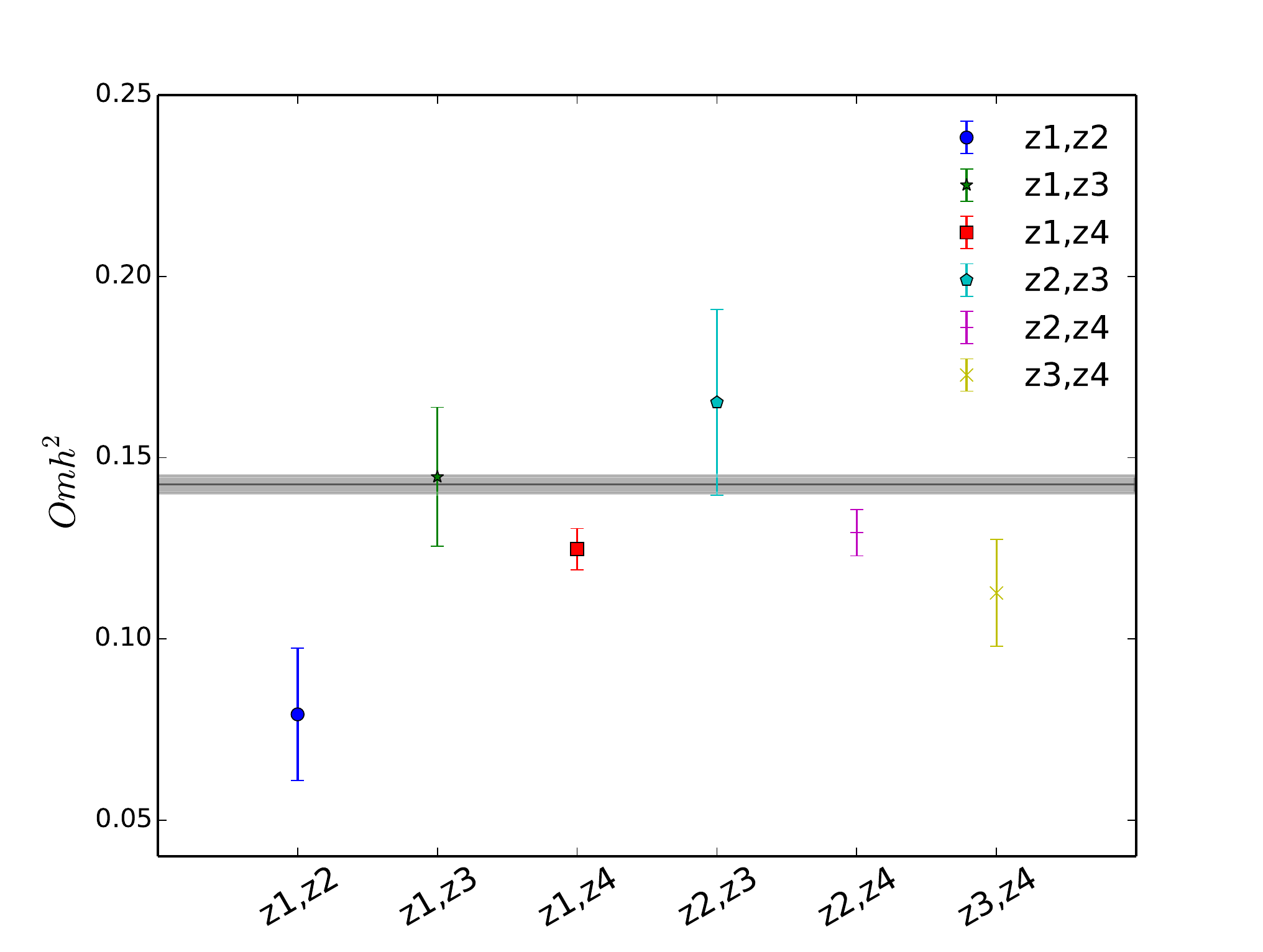}
    \caption{The six $Omh^2$ values computed from the four binned $H(z)$ data. Different colors refer to different pairs of 
    redshift bins. The uncertainty is estimated conforming to errors of each $H(z)$ pair. The black line and shaded region 
    correspond to the best-fit value of $\Om h^2$ and its uncertainties retrieved from the Planck CMB data.}
    \label{fig:omh2number}
\end{figure}

In $\Lambda$CDM, we have $Omh^2 \equiv \Om h^2$. The value of $\Om h^2$ is constrained tightly by the Planck observations to be 
centered around 0.14 for the base $\Lambda$CDM model fit\citep{2014A&A...571A..16P}: the Planck temperature power spectrum data 
alone gives $0.1423\pm0.0029$, the Planck temperature data with Planck lensing gives $0.1414\pm0.0029$, and the Planck temperature 
data with WMAP polarization at low multipoles gives $0.1426\pm0.0025$, all at $1\sigma$ confidence level (CL). In this paper, we 
choose $0.1426 \pm 0.0025$ as the Planck value. As shown in Figure \ref{fig:omh2number}, the testing pairs $Omh^2(z_{1};z_{3})$ 
and $Omh^2(z_{2};z_{3})$ are consistent with $\Lambda$CDM at $1\sigma$ CL, whereas for $Omh^2(z_{2};z_{4})$ and 
$Omh^2(z_{3};z_{4})$, $\Lambda$CDM can only be accommodated at $2\sigma$ CL. Particularly, for $Omh^2(z_{1};z_{2})$ and 
$Omh^2(z_{1};z_{4})$, $\Lambda$CDM is not compatible even at $2\sigma$ CL. Although deviations from $\Lambda$CDM exist for some 
pairs, cautiously, we cannot rule out the validity of $\Lambda$CDM.

\subsection{Measurements of $H_0$}\label{ssect:driveH0}

\begin{figure}
    \includegraphics[width=0.5\textwidth]{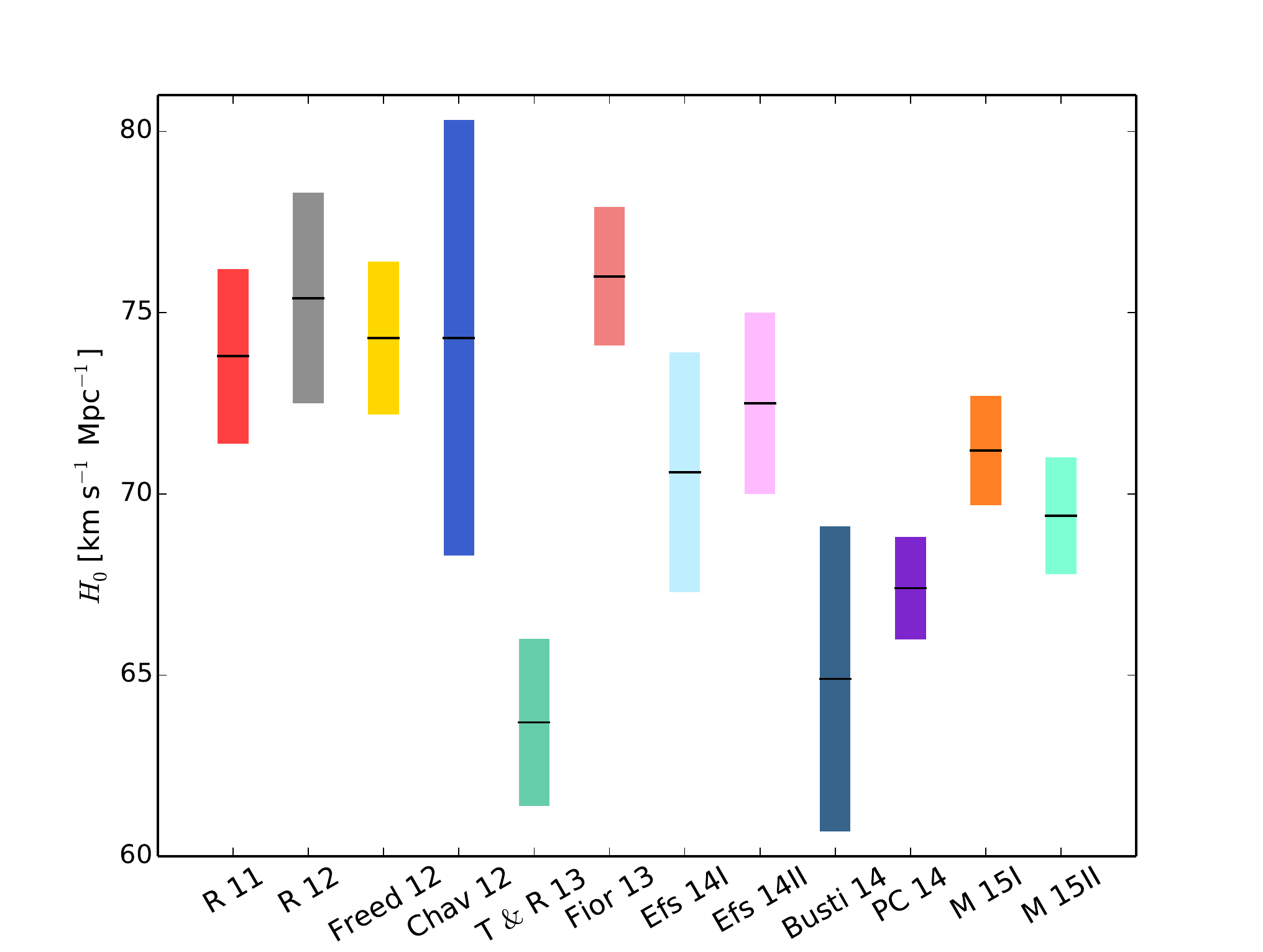}
    \caption{Different determinations of $H_0$ from different groups calculated using local calibrations or global parameter fits. 
    The first ten results, from left to right, represent the studies of  \citet{2011ApJ...730..119R}, \citet{2012ApJ...745..156R}, 
    \citet{2012ApJ...758...24F}, \citet{2012MNRAS.425L..56C}, \citet{2013A&A...549A.136T}, \citet{2013MNRAS.434.2866F}, 
    \citet{2014MNRAS.440.1138E} (with one distance anchor), \citet{2014MNRAS.440.1138E} (with three distance anchors), 
    \citet{2014MNRAS.441L..11B}, and \citet{2014A&A...571A..16P}, respectively. The last two values are the results from method 
    (I) and (II) presented in this work.}
    \label{fig:H0diffgroup}
\end{figure}

Recently, $H_0$ determination remains in disagreement about the range of possible values. Some measurements of $H_0$ are based on 
the calibrations from local distance indicators 
\citep{2011ApJ...730..119R,2012ApJ...745..156R,2012ApJ...758...24F,2012MNRAS.425L..56C,2013A&A...549A.136T,2013MNRAS.434.2866F,2014MNRAS.440.1138E}.  
Others come from the global parameter fit to CMB anisotropy observations \citep{2014A&A...571A..16P} and extrapolation
using existing $H(z)$ data \citep{2014MNRAS.441L..11B}. Some of the these results are also summarized in Figure \ref{fig:H0diffgroup}. 

Unlike the previous subsection, here we derive $H_0$ constraints utilizing the two-point $Omh^2(z_{2};z_{1})$ diagnostic, under 
the assumption that the values of $Omh^2(z_{2};z_{1})$ are the same for every $H(z)$ pair, namely, $\Lambda$CDM being
considered. The expression of the $Omh^2(z_{2};z_{1})$ diagnostic is thus modified into
\begin{equation}
   \label{eq:Omh2}
   Omh^2(z;0)=\frac{(\frac{H(z)}{100})^2-(\frac{H_{0}}{100})^2}{(1+z)^3-1}.
\end{equation}

\begin{figure}
    \includegraphics[width=0.5\textwidth]{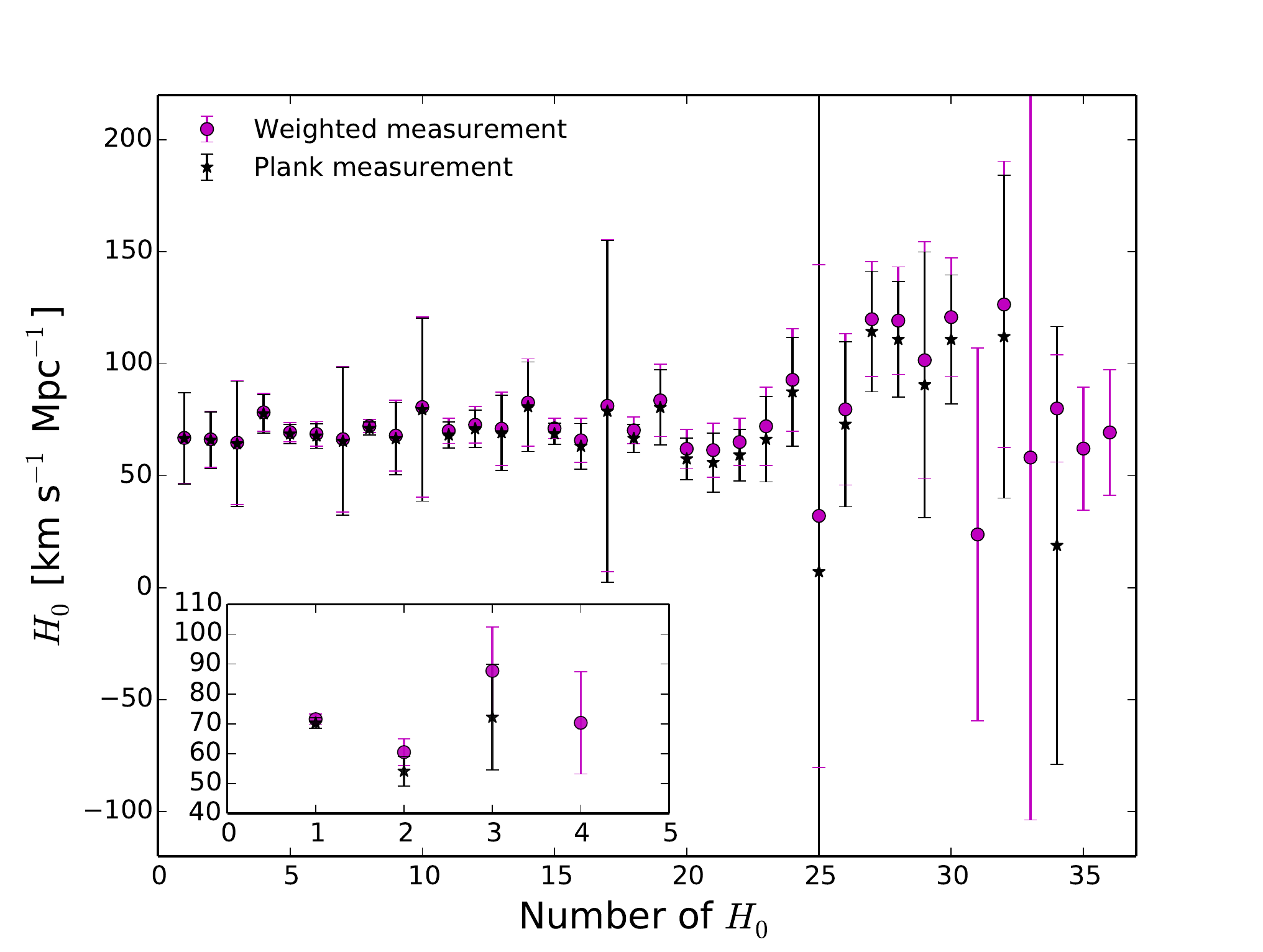}
    \caption{$H_0$ values from the binned and observational $H(z)$ data. Magenta points represent those derived from the inverse 
    variance weighted method; meanwhile, black stars mark the results from the Planck measurement. The values of $H_0$, from left 
    to right, are obtained from the thirty-six $H(z)$ data points in ascending order of redshift. The insert in this figure shows 
    derived $H_0$ values from binned $H(z)$ data with the same color coding.}
    \label{fig:H0number}
\end{figure}

A noteworthy characteristic of these $Omh^2$ diagnostic is that it shall be a constant in $\Lambda$CDM. In order to determine this 
constant, similar to what we have done in the previous subsection, we take an inverse variance weighted average of the six $Omh^2$ 
values from the four binned $H(z)$ points to get the final best-fit value and its uncertainty, i.e., $\overline{Omh^2} = 0.125 \pm 
0.004$. Combining with Equation \ref{eq:Omh2}, it is then straightforward to derive the value of $H_0$ (method (I)). In the meanwhile, 
we perform the same process except that the $Omh^2$ value originates from the Planck measurement (method (II)). The results 
given by these two approaches are shown in Figure \ref{fig:H0number} for comparison. Generally speaking, the values of $H_{0}$ 
from the Planck measurements are lower than those from our weighted method.

Utilizing methods (I) and (II) described above, we are able to determine all corresponding $H_0$ values from each $H(z)$ point 
(see Figure \ref{fig:H0number}). However, in method (II), there are four observational $H(z)$ data points which cannot be used to 
attain $H_0$ value, due to the minus values of $(\frac{H(z)}{100})^2-Omh^2(z;0)((1+z)^3-1)$. The associated redshifts of them are 
$z = 1.53$, $z = 1.965$, $z = 2.34$, and $z = 2.36$. We struggle with the same problem when evaluating $H_0$ from the fourth 
binned $H(z)$ point. One cause may be ascribed to systematics induced from observations and reduction procedures of both the 
$H(z)$ data and the Planck measurement; there are probably other reasons latent in deep explanations, such as the deviation of $\Lambda$CDM.

Then, we use the thirty-six $H_0$ measurements (from method (I)) and thirty-two $H_0$ measurements (from method (II)) to get 
the final weighted $H_0$ values. Figure \ref{fig:H0diffgroup} summarizes the $H_0$ results reported by different research groups. We 
obtain $H_0 = 71.23\pm1.54$ $\hunit$ in method (I) and $H_0 = 69.37\pm1.59$ $\hunit$ in method (II), both at $1\sigma$ CL (see the last two 
results in Figure \ref{fig:H0diffgroup}). Note that here the $H_0$ values from the two methods are derived assuming an underlying 
$\Lambda$CDM model.

\section{Uncertainty on DE Reconstruction from error of OHD}\label{sect:errorseffect}

One of the key issues in cosmology is that whether the DE evolving its EOS or not, which reveals the underlying properties of DE. The test of the deviation of the observational data from $\Lambda$CDM , as well as $w(z)$ evolution with respect to redshift is difficult. One way to overcome this problem is to directly reconstruct the effective DE EOS via observations, in which errors need to be considered attentively. The section above indicated a way to test $\Lambda$CDM with OHD. In this section, we try to explore how the error in OHD propagate into $w(z)$ at certain redshift during the reconstruction of $w(z)$.

A flat cosmological model with a DE term $\Omega_{\Lambda} = 1-\Omega_{m}$ is implemented to analyze the error propagation. We first discard the errors in parameters but $H(z)$ for simplicity. In this case, $w(z)$ and its uncertainty are reformulated as

\begin{equation}
\begin{aligned}
   \label{eq:wHz}
   w(z) 
   =& \frac{\log_{(1+z)}\left(\frac{H^2(z)}{H_{0}^2}-\Omega_{m}(1+z)^3\right)}{3} \\
   &-\frac{\log_{(1+z)}(1-\Omega_{m})}{3}-1,
\end{aligned}
\end{equation}

\begin{equation}
   \label{eq:deltawHz}
   \delta w(z) = \frac{2H(z)\delta H(z)}{3H_{0}^2\left(\frac{H^2(z)}{H_{0}^2}-\Omega_{m}(1+z)^3\right)\ln (1+z)}.
\end{equation}

\begin{figure}
    \includegraphics[width=0.5\textwidth]{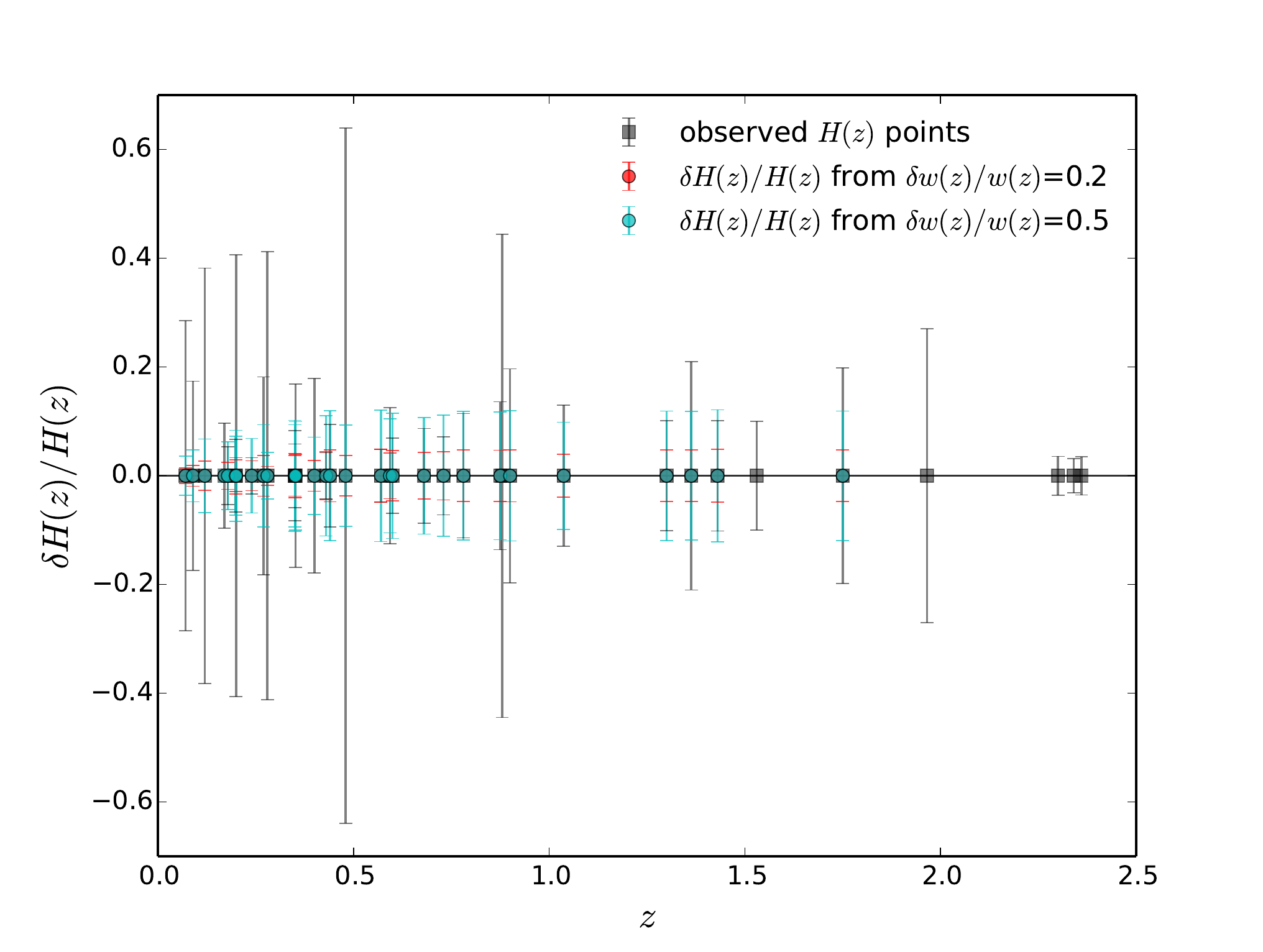}
    \caption{The ratio of required $H(z)$ error to the best fit value of OHD when the relative error of $w(z)$ is fixed at 20\% or 50\% with ideal assumption that the uncertainty of reconstructing $w(z)$ only comes from the prior on the error of $H(z)$. The gray error bar represents the ratio of observational error of $H(z)$ data to its best fit value. Red error bar means the ratio of predicted error of $H(z)$ data to its best fit value if $\delta w(z)/w(z)$ is fixed at 20\%. Cyan color represents the situation of $\delta w(z)/w(z)=50\%$.}
    \label{fig:deltaH_H}
\end{figure}

Equations \ref{eq:wHz} and \ref{eq:deltawHz} indicate the relation between $\delta H(z)$ and $\delta w(z)$. Here we fixed the relative errors of $w(z)$ to be 20\% and 50\% arbitrarily to derive the required $\delta H(z)$ to compare with the observed $H(z)$ errors. The results are shown in figure \ref{fig:deltaH_H}, where $H_0 = 67.1$ $\hunit$ and $\Omega_{m} = 0.3175$ are taken from the latest Planck constrains \citep{2014A&A...571A..16P}. It is worth noticing that high accuracy OHD is especially necessary in $w(z)$ reconstruction at low redshift. As an example, the error of $H(z)$ at $z=0.0708$ which is the nearest available OHD with the best fit value 69.0 $\hunit$, should be less than $\pm1.0$ $\hunit$ and $\pm2.5$ $\hunit$ if we expect a reconstruction of w(z) with accuracy within 20\% and  50\% respectively. It should be noted, though, that there are six OHD points which cannot be applied in w(z) reconstruction, due to the minus values of $\frac{H^2(z)}{H^2_0}-\Omega_{m}(1+z)^3$ (that means $\Omega_{m}>1$). The respective redshifts of them are $z = 0.88$, $z = 1.53$, $z = 1.965$, $z = 2.3$, $z = 2.34$, and $z = 2.36$.

However, the absence of errors in other parameters is of course an idealization. We then consider the more complex situation that the uncertainty of $w(z)$ not only come from the error of $H(z)$, but also have a bearing on the errors of $H_0$ and $\Omega_{m}$. What deserves special mention is that we do not consider the correlation between parameters owing to the lack of necessary information. In this case, the uncertainty of $w(z)$ is given by

\begin{equation}\label{deltawallpara}
\begin{aligned}
\delta w(z)
= & \Bigg[\Bigg(\frac{2H(z)}{3H^2_{0}\left(\frac{H^2(z)}{H^2_0}-\Omega_{m}(1+z)^3\right)\ln (1+z)}\Bigg)^2\delta^2H(z) \\
&+ \Bigg(\frac{-2H^2(z)H^{-3}_0}{3\left(\frac{H^2(z)}{H^2_0}-\Omega_{m}(1+z)^3\right)\ln (1+z)}\Bigg)^2\delta^2H_0 \\
&+ \Bigg(\frac{(1+z)^3}{3\left(\frac{H^2(z)}{H^2_0}-\Omega_{m}(1+z)^3\right)\ln (1+z)} \\
&- \frac{1}{3(1-\Omega_{m})\ln (1+z)}\Bigg)^2 \delta^2\Omega_{m}\Bigg]^\frac{1}{2} .
\end{aligned}
\end{equation}

\begin{figure}
    \includegraphics[width=0.5\textwidth]{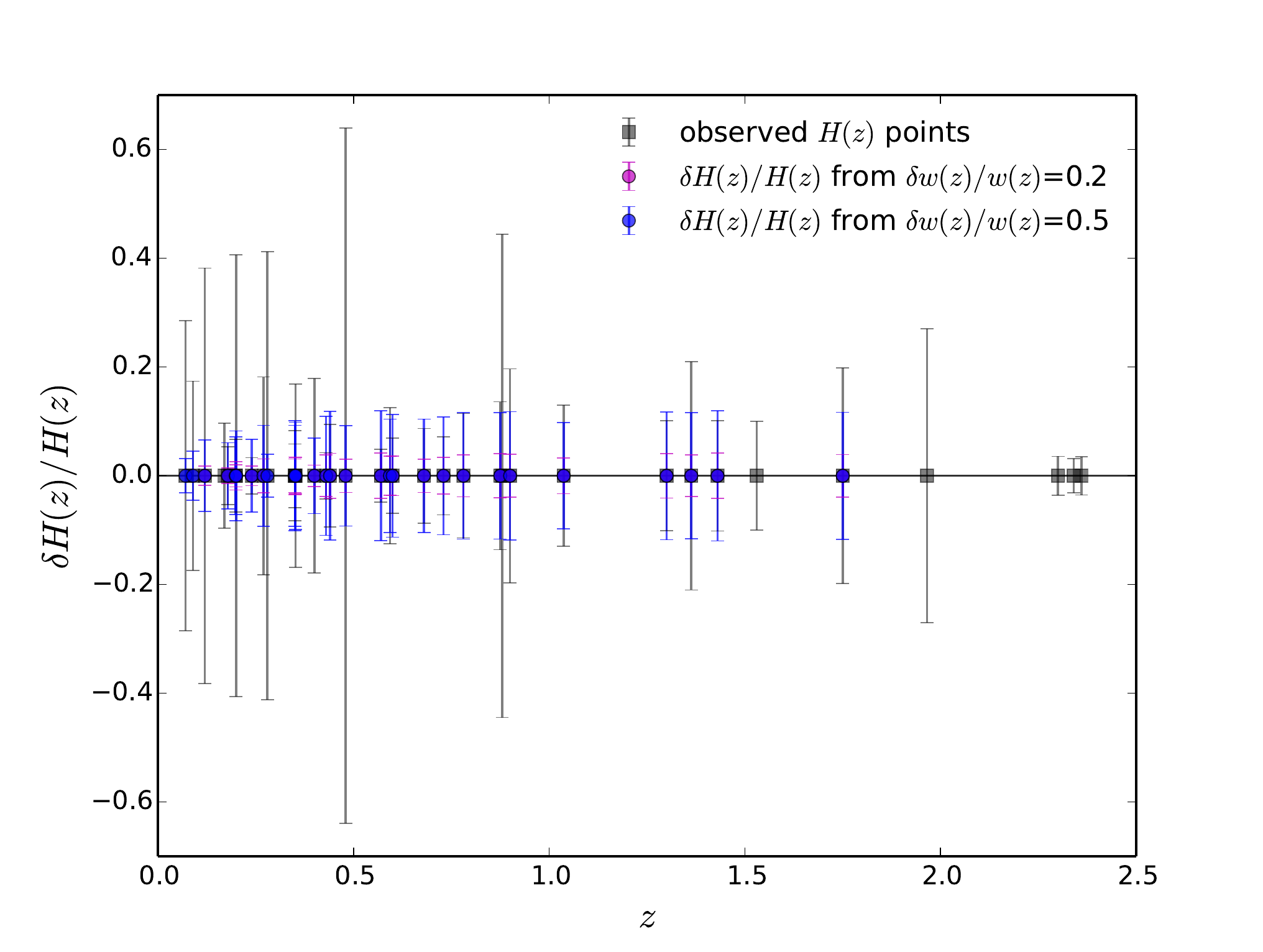}
    \caption{Same as Figure 4, except that the uncertainty on the reconstructed $w(z)$ come from the priors on the errors of $H(z)$, $H_0$, and $\Omega_{m}$. Magenta error bar means the ratio of predicted error of $H(z)$ to its best fit value if the uncertainty of $w(z)$ is fixed at 20\%. Blue color represents the situation of 50\%.}
    \label{fig:deltaH_H_allpara}
\end{figure}

Here we choose $H_0 = 67.4\pm1.4$ $\hunit$ and $\densm = 0.314\pm0.02$ as the prior which are constrained tightly by the Planck observation \citep{2014A&A...571A..16P}. Obviously, increased accuracy OHD is required if more parameters are considered to contribute to the uncertainty of $w(z)$. Reconsider the case of $H(z)$ at z=0.0708, when the relative uncertainty of $w(z)$ is constrained at  50\%, we could derive the required $\delta H(z)=\pm2.2 $ $\hunit$, however, required $\delta H(z)$ cannot be derived when $\delta w(z) / w(z)$ is fixed at 20\% because the relative uncertainty of $w(z)$ which comes from the errors of $H_0$ and $\Omega_{m}$ is already over 20\%. Some data points in figure \ref{fig:deltaH_H_allpara} do not have a required error bar (at $z = 0.0708$, $0.09$, $0.17$, and $0.28$ for $\delta w(z) / w(z)=20\%$; $z = 0.17$ for $\delta w(z) / w(z)=50\%$) comparing with that of in figure \ref{fig:deltaH_H} due to the same reason.

\section{Conclusions and Discussions} \label{sect:conclu}
In this paper, motivated by the investigation on the nature of DE, we have tested the validity of $\Lambda$CDM by the two-point $Omh^2(z_{2};z_{1})$ diagnostic using thirty-six OHD which are obtained from differential age and BAO methods. In order to obtain more reliable result, we binned all the OHD into four data points according to the redshift distribution to mitigate the observational contaminations. Moreover, in order to determine the best fit value and the uncertainty of binned $H(z)$, we took an inverse variance weighted average method. Our result furthers a discussion about the essence of DE, namely, we cannot rule out the validity of $\Lambda$CDM relying on nothing more than the current OHD. 

In a natural way, on the premise of $\Lambda$CDM, we further used the two-point $Omh^2(z_{2};z_{1})$ diagnostic to derive the Hubble parameter at $z=0$. For more comprehensive analysis, we adopted two methods to determine the value of $Omh^2$, one derived from binned $H(z)$ points (method (I)), the other was based on the result obtained from Planck measurement (method (II)). Throughout above two methods, corresponding $H_{0}$ value has been reconstructed from each observational $H(z)$ point logically. However, there are four OHD at $z = 1.53$, $z = 1.965$, $z = 2.34$, $z = 2.36$ cannot be used to attain the $H_{0}$ value by method (II). One reason may be attribute to systematics come from observations and reduction procedures; there can be others deep reasons, for example, the deviation of $\Lambda$CDM. Thus, attention should be paid to some thorough effort on improving spectra quality to yield more $H(z)$ determinations with small error in future work.

Utilizing the $H_0$ determinations with error corresponding to each $H(z)$ data derived by method (I) and (II), we reached the final weighted $H_0 = 71.23\pm1.54$ $\hunit$ (method (I)) and $H_0 = 69.37\pm1.59$ $\hunit$ (method (II)). Admittedly, our results about $H_0$ is tentative as well as model-dependent. The conclusion is obtained under the assumption of $\Lambda$CDM model. Therefore, care must be taken not extrapolating our conclusions beyond this assumption, especially when used in constraining cosmological parameters.

Furthermore, in the end of this paper we explored how the error of $H(z)$ impact on the uncertainty of $w(z)$. To illustrate the effect from $H(z)$ data, it is worth making a comparison between the two assumptions, namely, the uncertainty on $w(z)$ simply comes from the error of $H(z)$, or also involves the contributions from other parameters's uncertainties. We argue that the current precision on OHD is not sufficient small to ensure the reconstruction of $w(z)$ in an acceptable error range, especially at the low redshift. In other words, the ability of available individual OHD on the DE evolutionary study is very limited.

Finally, we note that it is necessary to place our main emphasis of research on BAO observation. Using BAO peak position as a standard ruler in the radial direction or combining measurements of BAO peak and Alcock-Paczynski distortion, we can limit the precision of $H(z)$ data better than 7\% \citep{2009MNRAS.399.1663G,2012MNRAS.425..405B}. As putting into operation of future space and ground-based telescopes (James Webb Space Telescope, Wide-Field Infrared Survey Telescope, planned adaptive optics systems with Keck, Large Synoptic Survey Telescope, and Thirty Meter Telescope et al.), more high-redshift, high-accuracy $H(z)$ determinations from BAO observations will undoubtedly perform a very useful role in the future study of the DE.

\acknowledgments \noindent \emph{Acknowledgments}. Tong-Jie Zhang thanks
Professor Martin White, and  Eric V. Linder for their  hospitality during visiting
Departments of Physics and Astronomy, University of
California, Berkeley and Lawrence Berkeley National
Laboratory. This work was supported by the National Science Foundation of China (Grants No. 11173006), the Ministry of Science and Technology National Basic Science program (project 973) under grant No. 2012CB821804.

\clearpage

\bibliographystyle{apj}
\bibliography{Omh2}

\end{document}